\documentclass[iop,apjl]{emulateapj}

\begin{document}
\title{Towards Precision Photometry with Extremely Large Telescopes:\\
the Double Subgiant Branch of NGC 1851}

\author{P. Turri\altaffilmark{1}}
\affil{Department of Physics and Astronomy, University of Victoria,\\
3800 Finnerty Road, Victoria, BC V8P 5C2, Canada}
\email{turri@uvic.ca}

\author{A. W. McConnachie, P. B. Stetson}
\affil{Herzberg Astronomy and Astrophysics, National Research Council Canada\\
5071 West Saanich Road, Victoria, BC V9E 2E7, Canada}

\author{G. Fiorentino}
\affil{Osservatorio Astronomico di Bologna, Istituto Nazionale di Astrofisica\\
Via Ranzani 1, 40127 Bologna, Italy}

\author{D. R. Andersen, J.-P. V{\'e}ran}
\affil{Herzberg Astronomy and Astrophysics, National Research Council Canada\\
5071 West Saanich Road, Victoria, BC V9E 2E7, Canada}

\and
\author{G. Bono\altaffilmark{2}}
\affil{Dipartimento di Fisica, Universit{\`a} di Roma Tor Vergata,\\
Via della Ricerca Scientifica 1, 00133 Roma, Italy}

\altaffiltext{1}{Herzberg Astronomy and Astrophysics, National Research Council Canada, 5071 West Saanich Road, Victoria, BC V9E 2E7, Canada}
\altaffiltext{2}{Osservatorio Astronomico di Roma, Istituto Nazionale di Astrofisica, Via Frascati 33, 00040 Monte Porzio Catone (RM), Italy}

\begin{abstract}
The Extremely Large Telescopes currently under construction have a collecting area that is an order of magnitude larger than the present largest optical telescopes. For seeing-limited observations the performance will scale as the collecting area but, with the successful use of adaptive optics, for many applications it will scale as $D^4$ (where $D$ is the diameter of the primary mirror). Central to the success of the ELTs, therefore, is the successful use of multi-conjugate adaptive optics (MCAO) that applies a high degree correction over a field of view larger than the few arcseconds that limits classical adaptive optics systems. In this letter, we report on the analysis of crowded field images taken on the central region of the Galactic globular cluster NGC 1851 in K$_\mathrm{s}$ band using GeMS at the Gemini South telescope, the only science-grade MCAO system in operation. We use this cluster as a benchmark to verify the ability to achieve precise near-infrared photometry by presenting the deepest K$_\mathrm{s}$ photometry in crowded fields ever obtained from the ground. We construct a colour-magnitude diagram in combination with the F606W band from HST/ACS. As well as detecting the ``knee'' in the lower main sequence at K$_\mathrm{s}\simeq$20.5, we also detect the double subgiant branch of NGC 1851, that demonstrates the high photometric accuracy of GeMS in crowded fields.
\end{abstract}

\keywords{globular clusters: individual (\object{NGC 1851}) --- instrumentation: adaptive optics --- techniques: photometric}

\section{Introduction}

\begin{figure*}[t]
\plotone{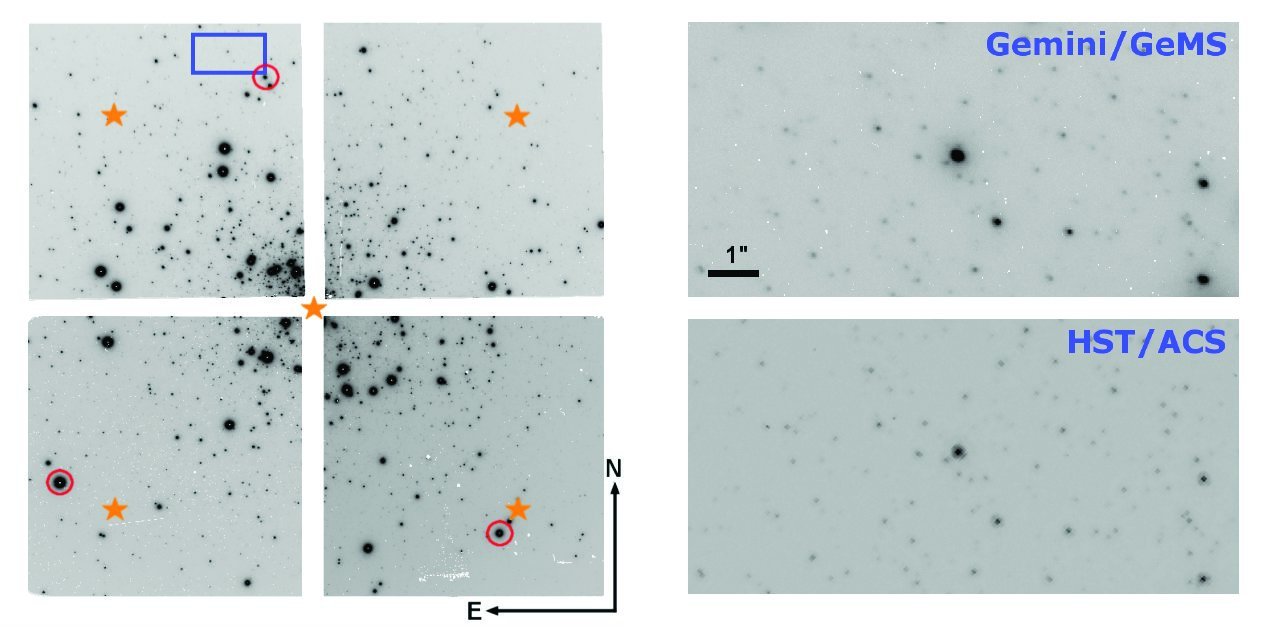}
\caption{\emph{Left:} Mosaic image of NGC 1851 taken with GeMS/GSAOI with a field of view of $83\arcsec{}\times 83\arcsec{}$. The positions of the five sodium LGSs are shown with orange markers and indicated by red circles are the three NGSs (clockwise from the left, $m_{R}=13.3,14.4,13.7$). \emph{Right:} a zoom-in of the region in the left panel contained within the blue rectangle observed with GSAOI in K$_\mathrm{s}$ band (\emph{top}) and the same region observed with HST/ACS in the optical by \cite{bib:sarajedini07} (\emph{bottom}). The spatial resolution of Gemini in the NIR and HST in the optical is comparable.\label{fig:mosaic}}
\end{figure*}

In classical adaptive optics (AO), the aberrated wavefront of a guide star is measured by a wavefront sensor and corrected by a deformable mirror (DM) to deliver a diffraction-limited image inside a field of view only several arcseconds wide, outside of which the point spread function (PSF) rapidly widens and deforms. This scale is set by the limiting angle inside of which the turbulence is correlated. A larger and more uniform field of view can be obtained with multi-conjugate adaptive optics (MCAO), by using an asterism of guide stars with multiple DMs optically conjugated at different heights of the atmosphere. Typically, AO works best in the near-infrared (NIR) because the spatial and temporal coherence of the aberrated wavefront is larger at longer wavelengths, resulting in better corrections (for a comprehensive review, see \cite{bib:davies12}).\\
The Gemini Multi-Conjugate Adaptive Optics System (GeMS) \citep{bib:rigaut14,bib:neichel14} on the Gemini South telescope on Cerro Pach{\'o}n in Chile is the first facility-class MCAO instrument in operation, although the concept has previously been proven with the Multi-Conjugate Adaptive Optics Demonstrator (MAD) on VLT \citep{bib:marchetti07}. GeMS uses two DMs and five sodium laser guide stars (LGS) to deliver a corrected field of view of $83\arcsec{}\times 83\arcsec{}$ to the Gemini South Adaptive Optics Imager (GSAOI) \citep{bib:mcgregor04,bib:carrasco12}, a mosaic of $2\times 2$ HAWAII-2RG chips with a pixel scale of 0.01962\arcsec{} px\textsuperscript{-1}.\\
Galactic globular clusters (GGCs) are an ideal target for MCAO systems. From an instrumentation perspective, their moderately large expanse on the sky and their high degree of crowding means that they are well-suited to MCAO systems that can potentially resolve faint stars in their central regions, while still possessing a large number of bright red natural guide stars (NGS) that are needed for the determination of the tip-tilt and defocus aberrations \citep{bib:rigaut92,bib:herriot06}. While deep and accurate NIR stellar photometry of GGCs has been a prerogative of the Hubble Space Telescope \citep{bib:milone12,bib:milone14}, in the future AO-assisted ELTs could get an edge over space telescopes in the NIR due to their much smaller diffraction limit.\\
From an astronomical perspective, the last decade has witnessed a revolution in our understanding of GCs. The current spectroscopic evidence indicates that all GGCs show star-to-star variations of C and N abundances together with well defined O-Na and Mg-Al anticorrelations in both evolved (RGB, HB) and unevolved stars. The fraction of GGCs showing also multiple sequences is limited and probably correlated with the total mass of the cluster \citep{bib:gratton12}. This confirmation of multiple stellar populations is in contrast to the classical view of a coeval cluster with homogeneous chemical abundances. In the case of NGC 1851, a double subgiant branch (SGB) has famously been observed by \cite{bib:milone08} using HST/ACS images. This feature has been interpreted as evidence of a second generation of star formation $\sim 1$ Gyr older than the first; alternatively, \cite{bib:cassisi08} suggested that the double SGB can be explained as two coeval populations, where one is a factor two more enhanced in CNO than the other.\\
Here, we present first results from a Gemini/GSAOI campaign targeting Galactic GCs using GeMS. In particular, we present observations of NGC 1851 taken in the K$_\mathrm{s}$ band that represent the deepest K$_\mathrm{s}$ photometry of a GC taken from the ground. We use this GC as a benchmark to demonstrate the precision of the MCAO data by producing a CMD of the stellar population with an HST/ACS optical band.

\section{Data}

\begin{figure*}[t]
\plottwo{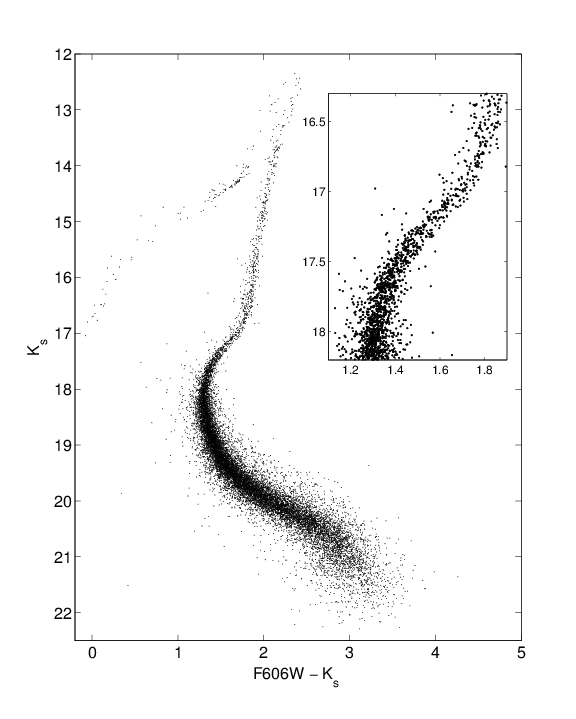}{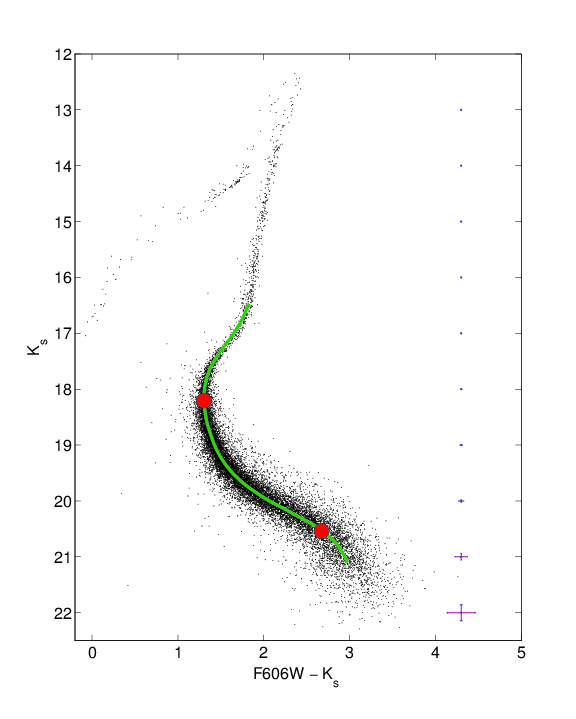}
\caption{\emph{Left:} (K$_\mathrm{s}$, F606W-K$_\mathrm{s}$) colour-magnitude diagram of NGC 1851; the detail of the double SGB is shown in the inset. \emph{Right:} same as the left panel with average photometric (random) uncertainties indicated. Overlaid is the fiducial line with the approximate locations of the main sequence turnoff and main sequence knee highlighted by red dots.\label{fig:cmd}}
\end{figure*}

During the GeMS/GSAOI Science Verification phase in Fall 2012, we obtained images of NGC 1851 in the J ($\lambda =1.235$ \micron) and K$_\mathrm{s}$ ($\lambda =2.159$ \micron) bands. Subsequent observing runs targeted five other globular clusters in the same bands. For this first paper in this series, we describe analysis of the K$_\mathrm{s}$ band data for NGC 1851; a full analysis of the J and K$_\mathrm{s}$ band discussing the properties of NGC 1851 will be published in a forthcoming article, and the other GCs will be discussed in subsequent papers. The observation with MCAO of globular clusters in NIR will prove particularly useful to study low mass stars, giving information on the total cluster mass, the lower end of the mass function and the colour-temperature relation for cold atmospheres. We will also discuss the use of these data for astrometric analyses.\\
The median seeing of our observations was recorded by RoboDIMM on Cerro Tololo to be of 0.77\arcsec{} and the instrument delivered in K$_\mathrm{s}$ band an average corrected full width at half maximum (FWHM) of 0.09\arcsec{}, comparable to the 0.06\arcsec{} FWHM of the diffraction limit case. The integration consisted of 12 dithered sub-exposures of 160 s each, and the time from the first to last science image was 75 minutes. Two short exposures of 21 s and 90 s were observed to obtain photometry for the brighter sources that were saturated in our longer exposures.\\
Image reduction has been performed using the Gemini GSAOI IRAF package that contains a bad pixel map and appropriate non-linearity correction. A super-flat was made that combined the twilight and dome flat fields, utilising the uniform spatial illumination of the former and the high signal-to-noise ratio of the latter. The low dark current of the H2RG detectors means we did not need to apply a dark frame. Local sky estimates were used for the point source photometry (see below), and so we did not need to apply a sky frame either. However, it is worth noting that there is a substantial number of bad pixels in each chip, and that one of the chips has a $\sim 0.2$ mag higher efficiency than the other. Figure~\ref{fig:mosaic} shows an example of a processed image.\\
The DAOPHOT suite of programs \citep{bib:stetson87,bib:stetson88,bib:stetson94} has been used to perform the photometric analysis on the individual frames, having already been proven to work with MCAO data \citep{bib:bono10,bib:fiorentino11,bib:davidge14,bib:saracino15}. To remove cosmic rays and bad pixels from our catalogue, we keep only objects that are found within a radius of one pixel in at least three dithered exposures out of twelve. Particular care has been paid to the PSF fitting routine due to the expected shape of the MCAO PSF. AO concentrates the light into the core of the PSF, but only up to a certain scale given by the density of actuators in the DM, beyond which the PSF remains uncorrected \citep{bib:jolissaint06}. Thus, even though the fraction of light in the core of the PSF is substantially greater than in the seeing-limited case, the size of the halo relative to the core is much larger in the AO case, and the adopted PSF model should reflect this. In addition, even although MCAO provides a more uniform correction over a larger field than classical AO, some significant spatial and temporal variation in the PSF structure may still be expected. Therefore, each exposure has $\sim 300$ manually selected bright and isolated stars used to model the PSF as an analytic profile plus a spatially-varying look-up table, where each pixel of the PSF is allowed to spatially vary as a cubic function, independent of every other pixel of the PSF. While we conduct our analysis independently on every exposure of every chip, we also perform an analysis on the stacked median image of each chip in order to detect and photometer the faintest stars. We ensure a high fidelity for our stellar catalogue by cross-matching every detection to the catalogue of NGC 1851 from the ACS Survey of Galactic Globular Clusters \citep{bib:sarajedini07}. Noise features and artefacts that are incorrectly identified as stellar objects are removed by this process, while real stars are kept. The full HST/ACS catalogue is considerably deeper than the K$_\mathrm{s}$ catalogue (by $\sim$4.5 magnitudes), and so this does not limit the depth of the K$_\mathrm{s}$ band data presented.\\
Each frame of each chip is calibrated independently to the K$_\mathrm{s}$ band magnitude scale of the 2MASS Point Source Catalog through an intermediate catalogue generated by one of the authors (PBS) from 256 seeing-limited archival images obtained with CTIO/NEWFIRM. These latter exposures are both wide and shallow enough to contain a large number ($\sim 4000$) of stars present also in the 2MASS catalogue. These were used to calibrate the magnitudes of fainter stars in the same images, which could then be used to determine the photometric zero points of the GeMS images.

\begin{figure*}[t]
\plottwo{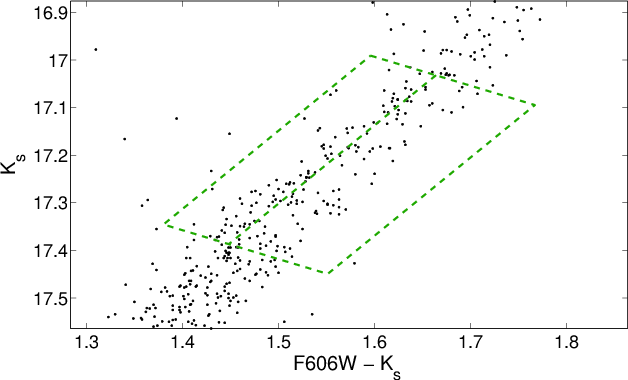}{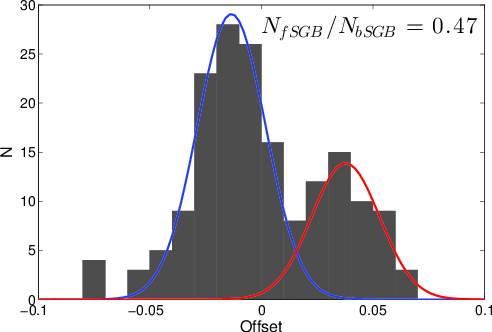}
\caption{\emph{Left:} As Figure~\ref{fig:cmd}, but enlarged in the region of the double SGB. Overlaid is the selection box orientated parallel with the linear segment of the SGB. \emph{Right:} histogram of the distribution of offsets relative to the median of the left panel box. Overlaid is the best-fit double Gaussian to the bimodal distribution.\label{fig:cmd_zoom}}
\end{figure*}

We note that the extended halos of the MCAO PSF, combined with the spatial variation in the shape of the PSF, can lead to a photometric zero-point that changes between sub-exposures. This will be discussed and explored in more detail in Turri et al. (2015, in preparation). The data presented in this letter have been corrected for this variable zero-point, but no color correction is applied, however we estimate it to be small. Note that this makes no difference to the analysis presented herein. For each star, the final magnitude was calculated by combining the magnitudes measured in each exposure using artificial skepticism \citep[chap.~3]{bib:stetson89}. The reddening correction in the direction of NGC 1851 is small ($E(B-V)=0.02$, \cite{bib:harris96}) and has not been applied.

\section{Results}
Figure~\ref{fig:cmd} shows the (K$_\mathrm{s}$, F606W-K$_\mathrm{s}$) CMD for NGC 1851 obtained by our analysis, and contains $\sim 16000$ stars.  The error bars are a combination of random uncertainty (photon and readout noise), PSF fitting residuals and scatter of the measurements between exposures. The F606W data are taken from \cite{bib:sarajedini07}, and the K$_\mathrm{s}$ data have additionally had quality control criteria in sharpness and $\chi^2$ applied. In addition, a coarse proper motion cut has been applied to reject foreground stars by restricting the apparent difference in the positions of the stars between the two epochs to 8 mas (the HST/ACS data were observed in 2006). More details and astrometric analysis will be presented in Turri et al. (2015, in preparation).\\
Several notable features are visible in Figure~\ref{fig:cmd}. Firstly, the extended horizontal branch is clearly visible and the red HB is particularly well defined. The asymptotic giant branch can be seen departing from the red HB just prior to the point where our data are saturated. In addition, \cite{bib:saviane98} have observed the red giant branch bump in NGC 1851 at a magnitude similar to the red HB. This feature occurs when the H-burning shell moving outwards approaches the chemical discontinuity caused by the first dredge-up of the convective envelope, and it can be observed in Figure~\ref{fig:cmd} on the RGB at K$_\mathrm{s}$=13.8.\\
The depth of the K$_\mathrm{s}$ is of particular note. Figure~\ref{fig:cmd} reaches $\sim$3.5 magnitudes deeper than the main sequence turnoff with a magnitude error of 0.15 at K$_\mathrm{s}$=22 using a total exposure of 1920 s. This depth is impossible to achieve for a crowded target like NGC 1851 without the use of MCAO. A similar depth in K$_\mathrm{s}$ has been obtained by \cite{bib:monelli15} in M 15 using FLAO on LBT but $3\arcmin{}$ from the center of the cluster.\\
A bend in the main sequence is visible towards the limits of our data, and starts at K$_\mathrm{s}\simeq$ 20.5. This feature, known as the main sequence knee (MSK) is a typical feature in NIR colour-magnitude diagrams of GCs and is caused by collisionally induced absorption of molecular hydrogen in cold stellar atmospheres \citep{bib:bono10,bib:monelli15}. It has been proposed that the relative distance between the MSK and the main sequence turnoff can be used as a distance and reddening independent cluster age determination \citep{bib:calamida09}, being the position of the former point independent of age. We will revisit this claim in future papers in this series.\\
Of particular interest, the subgiant branch in Figure~\ref{fig:cmd_zoom} is clearly split into two sequences. This feature was first identified for NGC 1851 by \cite{bib:milone08} using HST/ACS data. Previous observations of the core of a GC with double SGBs from the ground have revealed at most a broadened SGB with a bimodal distribution (specifically, the excellent study of NGC 6656 by \cite{bib:libralato14}); only observations of the uncrowded outskirts have been able to discern multiple SGBs from seeing-limited images \citep{bib:milone09,bib:han09,bib:cummings14}. The fact that this feature is visible in the centre of NGC 1851 using the K$_\mathrm{s}$ band data from GeMS is a verification of our small photometric uncertainties which demonstrates significant promise for the successful exploitation of MCAO for precision photometry of crowded fields in the ELT era.\\
The right panel of Figure~\ref{fig:cmd_zoom} shows a histogram of the offsets of the stars contained within the box in the left panel. Two separate peaks are clearly identifiable and these correspond to the faint and bright SGB (fSGB, bSGB) as described by \cite{bib:milone08}. We model this distribution as two Gaussian curves that are overlaid on the histogram. By integrating these curves, we find $N_{fSGB}/N_{bSGB}=0.47$, entirely consistent with the findings of \cite{bib:milone09}.

\section{Conclusion}
We present first results from a study of the core of NGC 1851 using the GeMS MCAO system on Gemini South. We demonstrate the ability of this system to achieve precise near-infrared photometry of crowded regions by presenting the deepest K$_\mathrm{s}$ observations ever obtained from the ground. In the colour-magnitude diagram created with the addition of the HST F606W band, we detect the ``knee'' in the lower main sequence and the double subgiant branch of the globular cluster. A subsequent paper will present a more detailed analysis of NGC 1851 using these data, as well as an analysis of the photometric and astrometric performance and stability of the GeMS system. Here, we have used this cluster as a benchmark of the quality of the MCAO photometry that has been obtained. This work is part of a larger project with GeMS to obtain J, K$_\mathrm{s}$ photometry for several GCs in the southern hemisphere some of which have NIR data from HST/WFC3, e.g. NGC 2808 \citep{bib:milone12}.\\
MCAO systems are an integral part of the two largest ELT projects as they tackle science programs that require an extended field of view, with NFIRAOS as a first light system for the Thirty Meter Telescope \citep{bib:boyer14}, and MAORY on the European-ELT \citep{bib:diolaiti14}. As the only science-grade MCAO system, GeMS is an important new facility instrument that can serve as a scientific testbed for developing a practical understanding of the challenges of obtaining accurate photometry and astrometry using such a system. This pilot study implies that high precision photometry can be achieved from the ground using MCAO, and can inform the development of appropriate data processing techniques and observing strategies to ensure the ELTs deliver their full scientific promise over extended fields of view.

\acknowledgments
Based on observations obtained at the Gemini Observatory, which is operated by the Association of Universities for Research in Astronomy, Inc., under a cooperative agreement with the NSF on behalf of the Gemini partnership: the National Science Foundation (United States), the National Research Council (Canada), CONICYT (Chile), the Australian Research Council (Australia), Minist\'{e}rio da Ci\^{e}ncia, Tecnologia e Inova\c{c}\~{a}o (Brazil) and Ministerio de Ciencia, Tecnolog\'{i}a e Innovaci\'{o}n Productiva (Argentina). Acquired through the Gemini Science Archive and processed using the Gemini IRAF package.\\
G. Fiorentino has been supported by the FIRB 2013 (MIUR grant RBFR13J716).\\
We appreciate the prompt support received from the GeMS staff for the image reduction. We are grateful to the adaptive optics group at NRC Herzberg for feedback on the system performance and to INAF - Osservatorio Astronomico di Bologna for insights on stellar populations in globular clusters.

{\it Facilities:} \facility{Gemini:South (GeMS, GSAOI)}.

\end{document}